\documentclass[%
 reprint,
 amsmath,amssymb,
 aps,
pra,
]{revtex4-1}

\usepackage{graphicx}% Include figure files
\graphicspath{ {images/} }
\usepackage{dcolumn}% Align table columns on decimal point
\usepackage{bm}% bold math
\usepackage{hyperref}% add hypertext capabilities
\usepackage{amssymb}
\usepackage{color}

\def\##1{{\bf #1}}
\def\=#1{\underline{\underline #1}}

\def\.{\mbox{ \tiny{$^\bullet$} }}

\def\le{\left(}
\def\ri{\right)}
\def\les{\left[}
\def\ris{\right]}

\def\epso{\epsilon_{0}}
\def\lambdao{\lambda_{ 0}}
\def\muo{\mu_{ 0}}
\def\ko{k_{ 0}}

\def\eps{\epsilon}

\def\ux{\hat{\#u}_x}
\def\uy{\hat{\#u}_y}
\def\uz{\hat{\#u}_z}

\newcommand{\dm}[1]{\underline{\underline{#1}}}

\usepackage{mathtools}

\begin{document}

%\preprint{APS/123-QED}

\title{Excitation of the Uller--Zenneck electromagnetic surface waves in the prism-coupled configuration}% Force line breaks with \\

\author{Mehran Rasheed}
% \altaffiliation[Also at ]{Department of Physics, Lahore University of Management Sciences (LUMS), Lahore, Pakistan.}%Lines break automatically or can be forced with \\
\author{Muhammad Faryad}%
 \email{muhammad.faryad@lums.edu.pk}
\affiliation{
 Department of Physics, Lahore University of Management Sciences (LUMS), Lahore, Pakistan.
}

\date{\today}% It is always \today, today,

\begin{abstract}
A configuration to excite the Uller--Zenneck surface electromagnetic waves at the  planar interfaces of homogeneous and isotropic dielectric materials is proposed and theoretically analyzed. The Uller--Zenneck waves are surface waves that can exist at the planar interface of two dissimilar  dielectric materials of which at least one is a lossy dielectric material. In this work, a slab of a lossy dielectric material was taken with lossless dielectric materials on both sides. A canonical boundary-value problem was set up and solved to find the possible Uller--Zenneck waves and waveguide modes. The Uller--Zenneck waves guided by the slab of the lossy dielectric material were found to be either symmetric or anti-symmetric that transmuted into waveguide modes when the thickness of that slab was increased. A prism-coupled configuration was then successfully devised to excite the Uller--Zenneck waves. The results showed that the Uller--Zenneck waves are excited at the same angle of incidence for any thickness of the slab of the lossy dielectric material, whereas the waveguide modes can be excited when the slab is sufficiently thick. The excitation of Uller--Zenneck waves at the planar interfaces with homogeneous and all-dielectric materials can usher new avenues for the applications for electromagnetic surface waves.
\end{abstract}

\maketitle

%\tableofcontents

\section{\label{sec:Intro}Introduction}
Surface electromagnetic waves are guided by an interface of two different materials and their power is localized to that interface. The most famous surface waves are surface plasmon-polariton waves (SPP) guided by the interface of a metal and a homogeneous dielectric material \cite{Raetherbook} and find applications in optical sensing, sub-wavelength imaging, and transmission of light through subwavelength holes \cite{JH2006,SAM2007}. However, much before the discovery of SPP waves, Uller in 1903 \cite{KU1903} and Zenneck in 1907 \cite{JZ1907}   have investigated the surface waves guided by the planar interface of two dielectric materials of which one is lossy. Whereas Uller consider the interface between dissipative sea water and air, Zenneck investigated the surface waves guided by air/ground interface. They were both interested in the long distance propagation of radio waves.  Electromagnetically, both problems are the same since one partnering material is almost lossless and the other is lossy.  The surface waves guided by the interface of a lossy and a lossless dielectric material are thus called Uller--Zenneck waves \cite{MF2014}.

The SPP waves and Uller--Zenneck waves are similar in the sense that both are $p$-polarized when both the partnering materials are homogeneous and isotropic. Furthermore, the expression of the relative wavenumber $q/\ko=\sqrt{\eps_s\eps_d/(\eps_s+\eps_d)}$ is the same for both, where $\eps_d$ is the permittivity of the lossless dielectric partner and $\eps_s$ is the permittivity of the metal (for SPP waves) or the lossy dielectric material (for the Uller--Zenneck waves) \cite{AL2013}. The major difference between the SPP waves and the Uller--Zenneck waves is that the magnitude of the real part of the relative wavenumber $q/\ko$ relative to the refractive index $\sqrt{\eps_d}$ of the lossless-dielectric-partner: For the SPP waves, the real part of the relative wavenumber is greater than the refractive index of the dielectric partner while for the Uller--Zenneck waves it is smaller, though by a small factor. Therefore, the phase speed of the Uller--Zenneck waves is greater than that in the bulk partnering lossless dielectric material, whereas the phase speed of the SPP wave is smaller than that in the bulk dielectric partner.

%%%%%%%%%%%%%%%%%%%%%%%%%%%%%%%%%%%%%%%%%
\begin{figure}
\includegraphics[width = 3in]{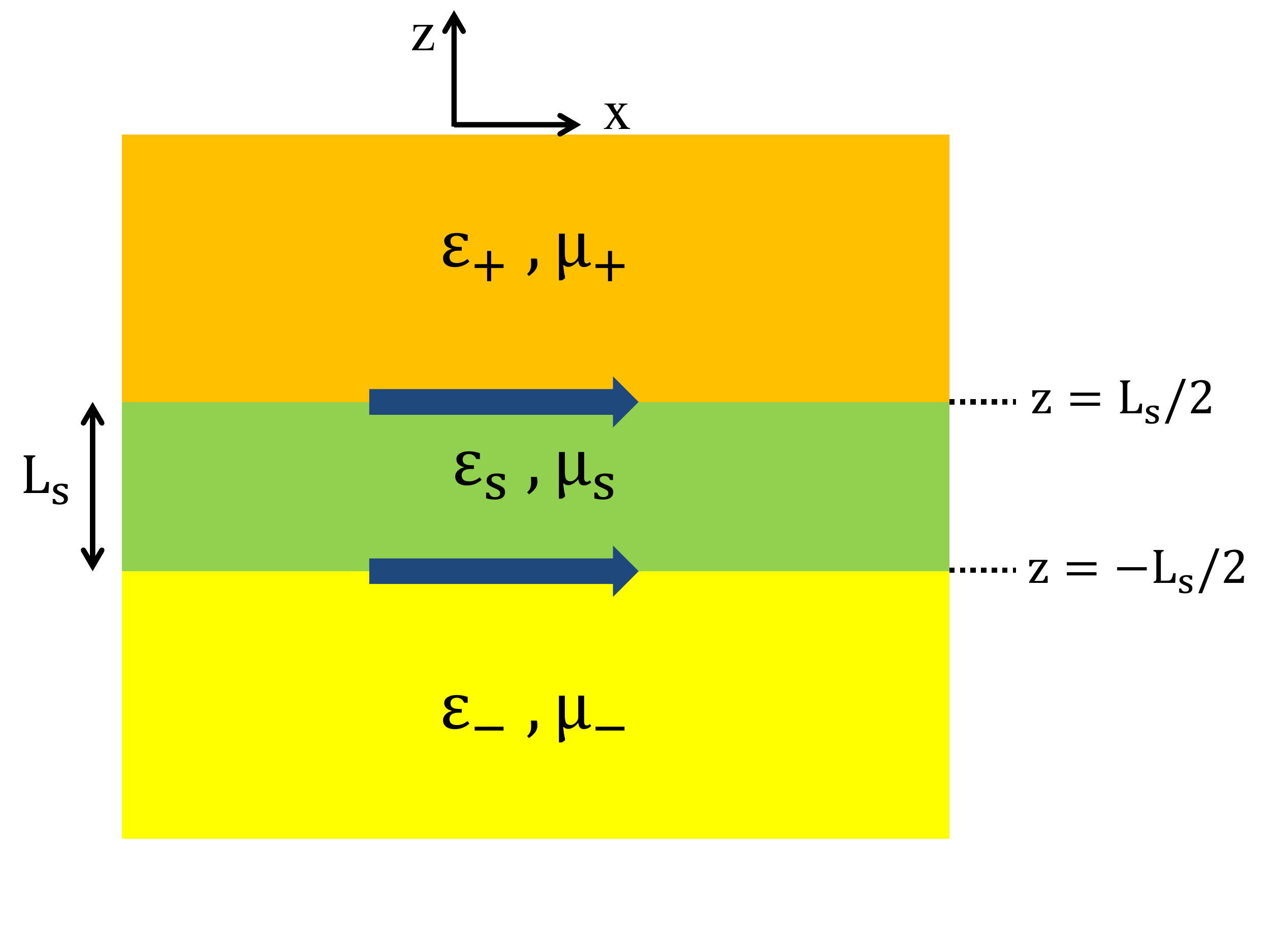}
\caption{{\bf Schematic of the canonical boundary-value problem:} The Uller--Zennck waves (horizontal, blue, thick arrows) are guided by a slab of lossy dielectric material with lossless dielectric materials occupying half spaces on either side.} 
\label{CanBP} 
\end{figure}
%%%%%%%%%%%%%%%%%%%%%%%%%%%%%%%%%%%%%%%%%%%

The solution of a canonical boundary-value problem of surface wave propagation by an interface between two semi-infinite expanses of a lossless and a lossy dielectric material clearly shows that the Uller--Zenneck waves can exist at a planar interface \cite{AL2013}. So, why were Uller--Zenneck waves left behind?  The major hurdle was the absence of a conclusive experimental proof that Uller--Zenneck waves can be excited, even though the theory was rigorously established by Sommerfeld after Uller and Zenneck  \cite{AS1909, AS1926, JRW1998}.  In the optical regime, it has recently been shown that the Uller--Zenneck waves could not be conclusively excited at the single planar interface of a lossless and a lossy dielectric material in a prism-coupled configuration \cite{MF2014}. However, it was theoretically shown that a surface-relief grating can excite the Uller--Zenneck waves \cite{MF2014}. Therefore, the first successful and unambiguous excitation of the Uller--Zenneck surface waves was accomplished  using a surface-relief grating in the optical regime \cite{MF2014_1}. However, the excitation on a single planar interface has not been successful so far, even though it is highly desirable to advance the scope of applications of the Uller--Zenneck waves since they offer a simplest way to excite surface waves.

Therefore, we set out to investigate the possibility of exciting the Uller--Zenneck waves at the planar interface using the coupling of the two planar interfaces in a prism-coupled configuration. For this purpose, a canonical boundary-value problem of surface-wave propagation by a thin slab of a lossy dielectric material was set up and solved to see if the Uller--Zenneck waves exist and can be differentiated from the possible waveguide modes that can also propagate in the planar slab of a lossy dielectric material. The formulation of the dispersion equation for the Uller--Zenneck waves guided by the slab of a lossy dielectric material and a brief description of the prism-coupled configuration is presented in Sec. \ref{theory}. The numerical results are presented and discussed in detail in Sec. \ref{sec:NumRes}. Finally, the concluding remarks are presented in Sec. \ref{conc}. A time dependence of an $\exp(-i\omega t)$ is assumed and suppressed throughout, where $i = \sqrt{-1}$, $\omega$ is the angular frequency, and $t$ is  the time. The free-space wavenumber and impedance is denoted by $\ko = \omega\sqrt{\varepsilon_{0}\mu_{0}}$ and $\eta_0=\sqrt{\muo/\epso}$, respectively,  where $\varepsilon_{0}$ is the permittivity of free space and $\mu_{0}$ is the permeability of free space. The vectors are denoted by bold symbols, column vectors are bold and placed in square brackets, and matrices are underlined twice and placed in square brackets. The Cartesian unit vectors are denoted as $\ux$, $\uy$, and $\uz$.

\section{Theoretical Formulation}\label{theory}
\subsection{\label{sec:CanBP_TF}Canonical Boundary-Value Problem}
The formulation of the canonical boundary-value problem for the single interface of a lossless and a lossy dielectric material is a textbook problem {\cite{Raetherbook,SAM2007,AL2013}}. The general formulation of a canonical problem of surface-wave propagation by a nonhomogeneous slab of a bianisotropic medium with different bianisotropic  mediums on either side has been provided in Ref. \cite{AL2013}.  A less general problem has been formulated in Refs. \cite{MF2011,MF2011_1}  where the surface-wave propagation by a slab of homogeneous material in a periodically nonhomogeneous anisotropic medium is presented. Here, we present the  formulation of the $p$-polarized Uller--Zenneck waves guided by a slab of lossy dielectric material with lossless dielectric materials on either side, all being isotropic.

Let us consider a homogeneous and isotropic but lossy dielectric slab of thickness $L_{s}$ with relative permittivity $\varepsilon_{s}$   between two half spaces of dielectric materials of relative permittivities $\varepsilon_{\pm}$ and  relative permeabilities $\mu_{\pm}$ in the half-spaces $z \gtrless \pm L_{s}/2$, as shown schematically in Fig. \ref{CanBP}.

Assuming the direction of propagation of the Uller--Zenneck waves to be $\ux$, the electric and magnetic field phasor  for the $p$-polarized wave can be written as
\begin{equation}
\begin{rcases}
&\boldsymbol{E}(\boldsymbol{r}) =\left [e_{x}(z)\ux + e_{z}(z)\uz\right] \exp{(iqx)} \\
&\boldsymbol{H}(\boldsymbol{r}) = h_{y}(z)\uy \exp{(iqx)}
\end{rcases}\,,
z \in [-\infty, \infty].
\end{equation}
The substitution of these field phasors in the Maxwell curl equations yield one algebraic and two first order differential equations. The algebraic equation gives
\begin{equation}
\begin{split}
e_z(z) = \frac{-q}{\omega \varepsilon_{0}\varepsilon(z)}\,, \qquad z \in [-\infty, \infty]\,,
\end{split}
\end{equation}
where
\begin{equation} \label{eq:eps_pmAndeps_s}
\begin{split}
 & \varepsilon(z)=\begin{cases}
               \varepsilon_{\pm}\,, \qquad z \gtrless \pm L_{s}/2\,,\\
               \varepsilon_{s}\,,  \qquad -L_{s}/2 < z <  L_{s}/2\,.
            \end{cases}
\end{split}
\end{equation}
The other two differential equations can be used to formulate the matrix ordinary differential equation
\begin{equation}
\frac{d}{dz}[\boldsymbol{f}^{(p)}(z)] = i[\dm{P}^{(p)}(z)]\cdot[\boldsymbol{f}^{(p)}(z)]\,, \label{MODE}\\
\end{equation}
where the $2\times1$ column vector 
\begin{equation}{\label{Eq_ppolfieldvec}}
[\boldsymbol{f}^{(p)}(z)]  = [e_{x}(z) \quad h_{y}(z)]^{T}\,,
\end{equation}
and the $2\times2$ square matrix 
\begin{equation} \label{eq:P_pmAndP_s}
\begin{split}
 & [\dm{P}^{(p)}(z)]=\begin{cases}
               [\dm{P}^{(p)}_{\pm}], \qquad z \gtrless \pm L_{s}/2\,,\\
               \\
               [\dm{P}^{(p)}_{s}],  \qquad -L_{s}/2 < z <  L_{s}/2\,,
            \end{cases}
\end{split}
\end{equation}
with
\begin{equation}
[\dm{P}^{(p)}_{\pm}] =
\begin{bmatrix}
0 & \frac{-q^2}{\omega\varepsilon_{0}\varepsilon_{\pm}} + \omega\mu_{0}\mu_{\pm} \\
\omega\varepsilon_{0}\varepsilon_{\pm}  & 0
\end{bmatrix}\,,
\end{equation}
and 
\begin{equation}
[\dm{P}^{(p)}_{s}] =
\begin{bmatrix}
0 & \frac{-q^2}{\omega\varepsilon_{0}\varepsilon_{s}} + \omega\mu_{0}\mu_{s} \\
\omega\varepsilon_{0}\varepsilon_{s}  & 0
\end{bmatrix}\,.
\end{equation}

For the surface waves, the field phasors in the half-spaces $z \gtrless \pm L_{s}/2$ must decay as $z \rightarrow \pm \infty$. In order to achieve this condition, only those eigenvectors of $[\dm{P}^{(p)}_{\pm}]$ have to be considered that represent decaying fields on either side.  The eigenvalues of  $[\dm{P}^{(p)}_{\pm}(z)]$ are found to be
\begin{equation}
\lambda^{(p)}_{(1)_{\pm}} =  \sqrt{k^2_{0}\varepsilon_{\pm}\mu_{\pm} - q^{2}}\,,
~\lambda^{(p)}_{(2)_{\pm}} = - \sqrt{k^2_{0}\varepsilon_{\pm}\mu_{\pm} - q^{2}}\,,
\end{equation}
with the corresponding eigenvectors
\begin{equation}
[\boldsymbol{t}^{(p)}_{1_{\pm}}] =
\begin{bmatrix}
1\\
\\
\frac{\omega \varepsilon_{0} \varepsilon_{\pm}}{\sqrt{k^{2}_{0}\varepsilon_{\pm}\mu_{\pm}-q^{2}}}
\end{bmatrix}\,,~[\boldsymbol{t}^{(p)}_{2_{\pm}}] =
\begin{bmatrix}
1\\
\\
-\frac{\omega \varepsilon_{0} \varepsilon_{\pm}}{\sqrt{k^{2}_{0}\varepsilon_{\pm}\mu_{\pm}-q^{2}}}
\end{bmatrix}\,.
\end{equation}

In the half-space $z > L_{s}/2$, $[\boldsymbol{t}^{(p)}_{1_{+}}]$ represent the decaying field since Im$\left[\lambda^{(p)}_{1_{+}}\right] > 0$. In the half-space $z < -L_{s}/2$, $[\boldsymbol{t}^{(p)}_{2_{-}}]$ represent the decaying fields since Im $\left[\lambda^{(p)}_{2_{-}}\right]< 0$. Therefore, we can write the fields at the boundaries on the sides of the half spaces as
\begin{equation}
  \bigg[\boldsymbol{f}^{(p)}\bigg(\frac{L_{s}}{2}+\bigg)\bigg] = a_{+}^{(p)}\big[\boldsymbol{t}_{1_{+}}^{(p)}\big]\,,\label{eq1}
\end{equation}
\begin{equation}
  \bigg[\boldsymbol{f}^{(p)}\bigg(-\frac{L_{s}}{2}-\bigg)\bigg] = a_{-}^{(p)}\big[\boldsymbol{t}_{2_{-}}^{(p)}\big]\,,\label{eq2}
\end{equation}
where $a_{\pm}^{(p)}$ are  unknown scalars. 

The solution of Eq. (\ref{MODE}) in the region $ -L_{s}/2 < z < L_{s}/2$ can be written as
\begin{equation}
\bigg[\boldsymbol{f}^{(p)}\bigg(\frac{L_{s}}{2}-\bigg)\bigg] = \exp\left\{i[\dm{P}_{s}]L_{s}\right\}\cdot\bigg[\boldsymbol{f}^{(p)}\bigg(-\frac{L_{s}}{2}+\bigg)\bigg]\,.\label{eq3}
\end{equation}
Implementing the standard boundary conditions $ [\boldsymbol{f}^{(p)}(L_{s}/2+)] =[\boldsymbol{f}^{(p)}({L_{s}}/{2}-)]$ and $[\boldsymbol{f}^{(p)}(-{L_{s}}/{2}+)]=[\boldsymbol{f}^{(p)}(-{L_{s}}/{2}-)]$ using Eqs. (\ref{eq1})--(\ref{eq3}), we get
\begin{equation}
\begin{split}
\begin{bmatrix}
[\boldsymbol{t}^{(p)}_{1_{+}}] \quad -\exp\left\{i[\dm{P}_{s}]L_{s}\right\}\boldsymbol{t}^{(p)}_{2_{-}}
\end{bmatrix}
 \cdot
\begin{bmatrix}
a^{(p)}_{+}\\
\\
a^{(p)}_{-}
\end{bmatrix}
= 0
\end{split}
\end{equation}
that can be written as
\begin{equation}
\begin{split}
\begin{bmatrix}
\dm{M}^{(p)}(q)
\end{bmatrix}
 \cdot
\begin{bmatrix}
a^{(p)}_{+}\\
\\
a^{(p)}_{-}
\end{bmatrix}
= 0
\end{split}\,.\label{Mamp}
\end{equation}
Therefore, the dispersion equation for the Uller--Zenneck waves is
\begin{equation}
\begin{split}
det
\begin{bmatrix}
\dm{M}^{(p)}(q)
\end{bmatrix}
= 0
\end{split}\label{dissp}
\end{equation}
so that the nontrivial solutions of Eq. (\ref{Mamp}) can exist.

\subsection{Prism-Coupled Configuration}\label{sec:PrismCoupConfig}
\begin{figure}
\includegraphics[width = 1\linewidth]{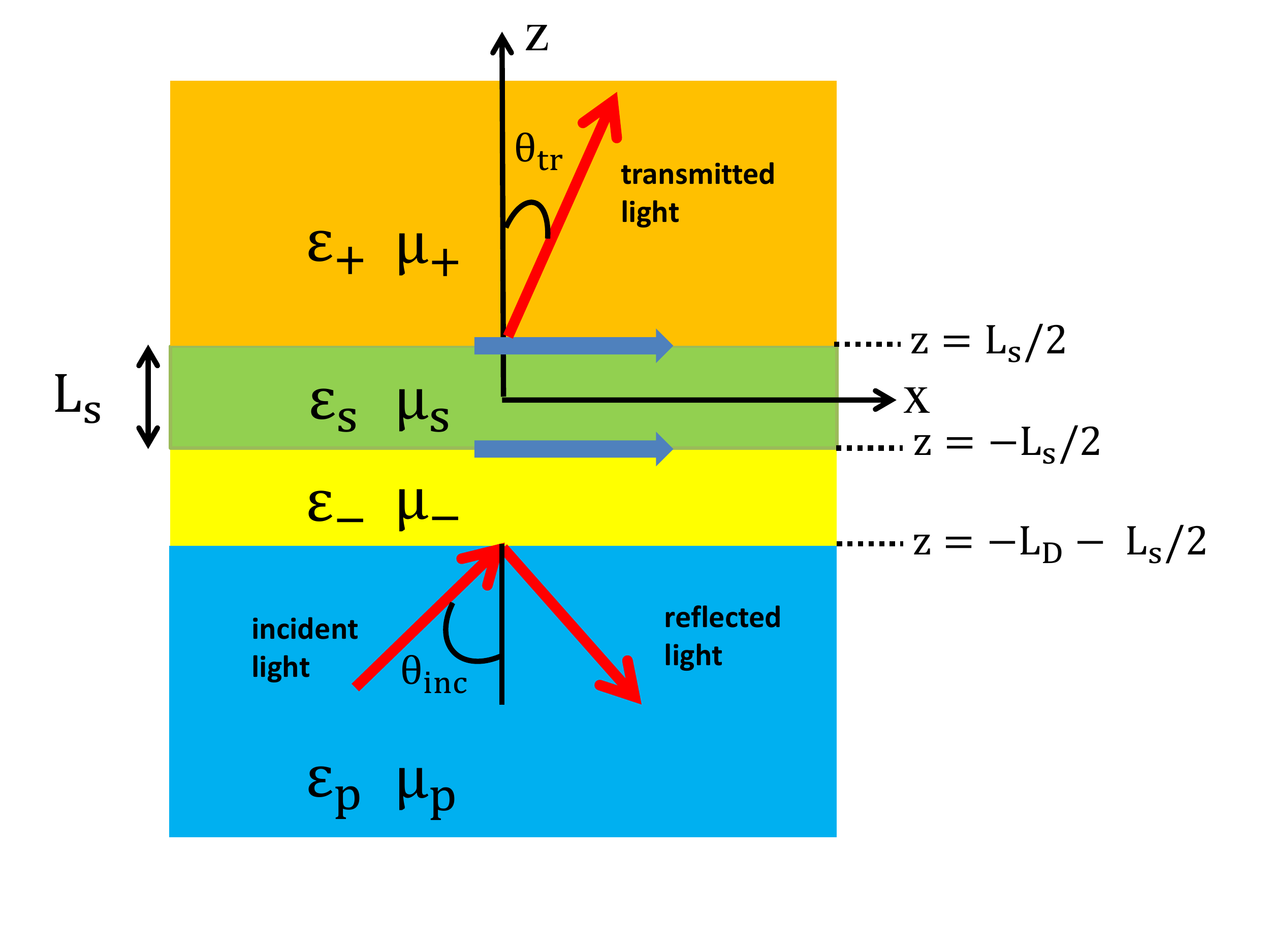}
\caption{{\bf Schematic of the prism-coupled configuration:} A $p$-polarized incident plane wave in the prism is incident on the finite layer of lossless dielectric layer sandwiched between the prism and the lossy dielectric slab of thickness $L_s$. The Uller--Zenneck waves (horizontal, blue, thick arrows) are excited at the boundaries of the lossy dielectric material.  }
\label{PCC}
\end{figure}

Let us now consider the prism-coupled configuration, schematically shown in Fig. \ref{PCC}. Only the description of the problem is presented as the formulation is straightforward and is available elsewhere \cite{AL2013}. The half-space $z <-L_D-L_s/2$ is assumed to be occupied by a prism with relative permittivity $\eps_p$ and permeability $\mu_p$. The region $ -L_D-L_s/2<z<-L_s/2$ is occupied by the lossless dielectric material with relative permittivity $\eps_-$ and relative permeability $\mu_-$ whereas the region $-L_s/2<z<L_s/2$ is occupied by the lossy dielectric material with relative permittivity $\eps_s$ and relative permeability $\mu_s$. The half-space $z >L_s/2$ is occupied by another lossless dielectric material with relative permittivity $\eps_+$ and relative permeability $\mu_+$.

%%%%%%%%%%%%%%%%%%%%%%%%%%%
  %%%%%%%%%%%%%%%%%%%%%%%%%

Let a $p$-polarized plane wave propagating in the prism ($z<-L_D-L_s/2$) be incident upon  the interface $z=-L_D-L_s/2$. The electric and magnetic field phasors of the incident plane wave in the prism can be written as 
\begin{equation} \label{eq:Inc_Field_PCC}
\begin{split}
&\boldsymbol{E}_{inc}(\boldsymbol{r}) = a_{p} (\ux \cos\theta_{inc} - \uz \sin\theta_{inc} )\\
 &\times \exp\left\{ ik_{0}n_{p}\left[x\sin\theta_{inc} + (z+L_D+L_s/2)\cos\theta_{inc}\right]\right\}\,,\\
 &\boldsymbol{H}_{inc}(\boldsymbol{r}) = a_{p} \frac{n_{p}}{\eta_{0}}\uy\\
 &\times \exp\left\{ ik_{0}n_{p}\left[x\sin\theta_{inc} + (z+L_D+L_s/2)\cos\theta_{inc}\right]\right\}\,,
 \end{split}
 \end{equation}
 where $n_p=\sqrt{\eps_p\mu_p}$ and $a_{p}$ is the amplitude of incident plane wave. 
The field phasors of the reflected plane waves can similarly be written as
\begin{equation} \label{eq:Ref_Field_PCC}
\begin{split}
&\boldsymbol{E}_{ref}(\boldsymbol{r}) = r_p(\ux \cos \theta_{inc} +\uz \sin \theta_{inc})\\
&\times \exp\left\{ ik_{0}n_{p}\left[x\sin\theta_{inc} - (z+L_D+L_s/2)\cos\theta_{inc}\right]\right\}\,,\\
&\boldsymbol{H}_{ref}(\boldsymbol{r}) = -r_{p} \frac{n_{p}}{\eta_{0}} \uy\\
 &\times \exp\left\{ ik_{0}n_{p}\left[x\sin\theta_{inc} - (z+L_D+L_s/2)\cos\theta_{inc}\right]\right\}\,,
\end{split}
\end{equation}
where $r_{p}$ is the amplitude of reflection.
Transmitted field phasors can be written as 
\begin{equation} \label{Tr_Field_PCC}
\begin{split}
&\boldsymbol{E}_{tr}(\boldsymbol{r}) = t_{p} (\ux \cos\theta_{tr} - \uz \sin\theta_{tr} )\\
 &\times \exp\left\{ ik_{0}n_{+}\left[x\sin\theta_{tr} + (z - L_s/2)\cos\theta_{tr}\right]\right\}\,,\\
 &\boldsymbol{H}_{tr}(\boldsymbol{r}) = t_{p} \frac{n_{+}}{\eta_{0}} \uy\\
 &\times \exp\left\{ ik_{0}n_{+}\left[x\sin\theta_{tr} + (z -L_s/2)\cos\theta_{tr}\right]\right\}\,,
 \end{split}
 \end{equation}
 where $n_+=\sqrt{\eps_+\mu_+}$ and $t_{p}$ is the amplitude of transmitted wave, and 
\begin{equation} \label{eq:Tr_Angle}
\sin \theta_{tr} = \frac{n_{p}}{n_{+}} \sin \theta_{inc}\,.
\end{equation}

The  reflectance and transmittance can be defined as
\begin{equation} \label{eq:Ref_Tr}
\begin{split}
&R_{pp} =\left | \frac{r_p}{a_p}\right | ^{2}\,,\\
&T_{pp} = \frac{n_{+}}{n_{p}} \frac{Re\{\cos \theta_{tr} \}}{\cos \theta_{inc}} \left|\frac{t_p}{a_p}\right|^2\,,
\end{split}
\end{equation}
and the absorptance can be computed using relation 
\begin{equation} \label{eq:Abs}
A_{p} = 1 - (R_{pp} + T_{pp})\,.
\end{equation}
Law of conservation of energy asserts that $A_{p}$ $\in$ $[0, 1]$.

\section{Numerical Results and Discussion}\label{sec:NumRes}
For all the numerical results, the free-space wavelength was fixed at $\lambdao=633$ nm. The lossy dielectric material  was taken to be silicon with $\varepsilon_{s} = 15.072 + 0.1521i$ and the mediums adjacent to the lossy material was taken to be silica $\eps_\pm=2.179$ \cite{RefInfo}. The prism was taken to be made of rutile with $\eps_p=n_p^2=(2.6)^2$. The relative permeability of all the media is taken to be the same and equal to $1$, i.e., $\mu_\pm = \mu_{s} = \mu_{p} = 1$.  

%%%%%%%%%%%%%%%%%%%%%%%%%%%%%%%%%%%%%%%%%%%%
\begin{figure}
\includegraphics[width = 1\linewidth]{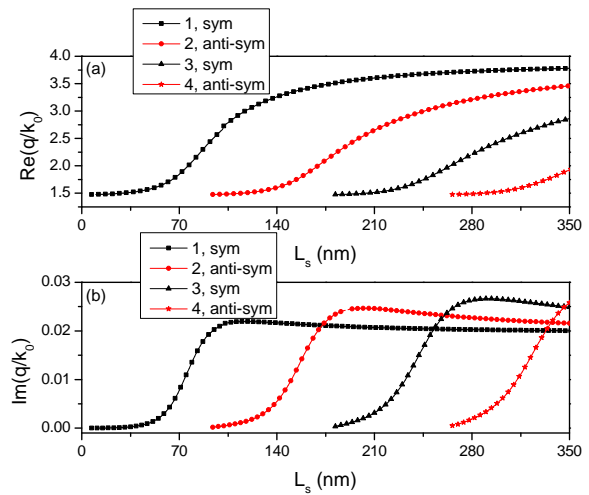}
\caption{{\bf Canonical problem:} Real and imaginary parts of relative wavenumbers $q/k_{0}$ of Uller--Zenneck waves  as a function of width $L_{s}$ of the lossy dielectric material, when $\lambda_{0} = 633$ nm, $\varepsilon_{\pm} = 2.179$, $\varepsilon_{s} = 15.072 + 0.1521i$, and $\mu_{\pm} = \mu_{s} = 1$. ``sym'' and ``anti-sym'' represent symmetric and anti-symmetric guided modes. All four lines in the graphs are for the same values of the parameters and represent multiple solutions of the dispersion equation (\ref{dissp}) at the same free-space wavelength $\lambda_0$.}
\label{qvsLplots}
\end{figure}
%%%%%%%%%%%%%%%%%%%%%%%%%%%%%%%%%%%%%%%%%%%%

%%%%%%%%%%%%%%%%%%%%%%%%%%%%%%%%%%%%%%%%%
\begin{figure*}
\includegraphics[width = 5.5in]{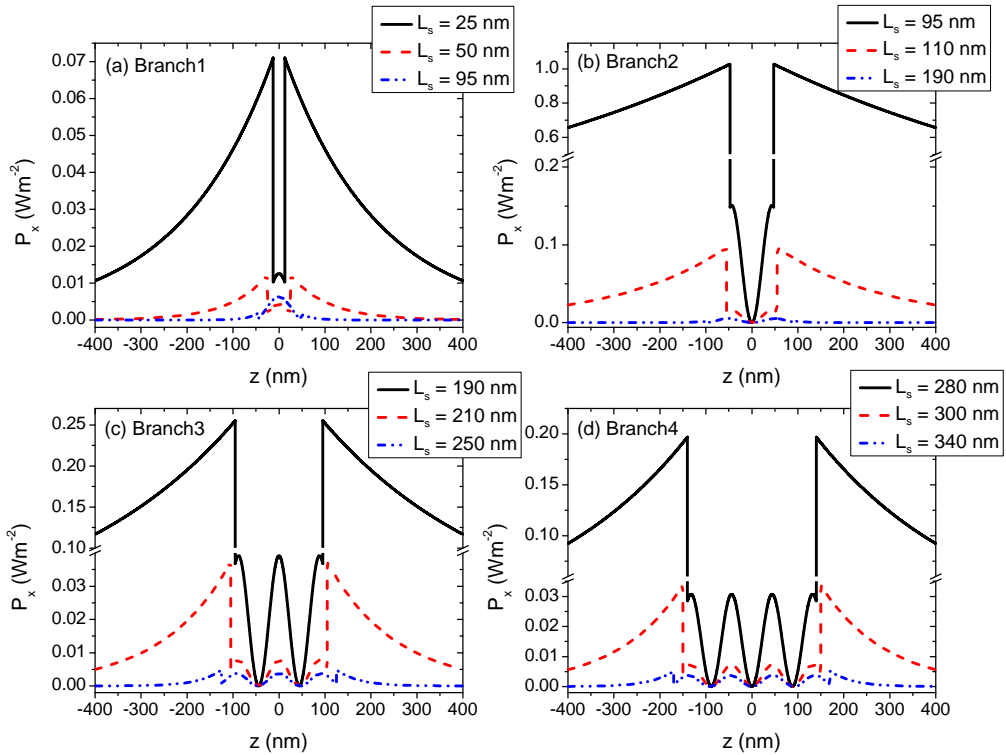}
\caption{{\bf Canonical problem:} The variation of the $x$-component of the time-averaged Poynting vector as a function of $z$ for three solutions on each of four branches in Fig. \ref{qvsLplots}. Other parameters are the same as for Fig. \ref{qvsLplots}. See Sec. \ref{canr} for the values of $q/\ko$ of each plot.} For the computations, $a_+^{(p)}=1$ Vm$^{-1}$ was set.
\label{CanPowPlotAll}
\end{figure*}
%%%%%%%%%%%%%%%%%%%%%%%%%%%%%%%%%%%%%%%%

\subsection{Canonical Problem}\label{canr}

The dispersion equation (\ref{dissp}) for the canonical problem was solved using the Newton--Raphson method and implemented in Matlab. The real and imaginary parts of the relative wavenumber $q/\ko$ as a function of $L_{s}\in[0,350]$ nm are presented  in the Fig. \ref{qvsLplots}. The figure shows that four solution branches are present, each starting at Re$(q/k_{0})\approx1.478$ and increasing  to Re$(q/k_{0})\approx3.685$ (when seen beyond $L_s=350$ nm). The first solution appears when $L_s\approx7$ nm.  For $L_{s} > 350$ nm (not shown in the figure) the similar trend was noted where new branches would similarly appear. Let us note that the intersections of branches representing the imaginary parts in Fig. \ref{qvsLplots}(b) are incidental and only indicate that the imaginary parts of the two solutions at these value of $L_s$ are the same. The real parts of the wavenumber at the intersection points are quite different as is apparent from Fig. \ref{qvsLplots}(a).  Our hypothesis is that each solution branch represents the Uller--Zenneck surface waves for the smaller values of $L_s$ and represents waveguide modes larger values of $L_s$. This is based on two main observations: (i) The relative wavenumber of the Uller--Zenneck waves for the {\textit single} interface $q/\ko=\sqrt{\eps_s\eps_\pm/(\eps_s+\eps_\pm)}=1.380+0.000879i$ is very close to the solutions of the dispersion equation for each branch for smaller values of $L_s$, and (ii) the spatial distribution of the time-averaged power, as discussed in the following paragraph.
 
This hypothesis is  confirmed by the examination of the spatial distribution of the $x$-component of the time-averaged Poynting vector 
\begin{equation}
P_{x}(z)  = -\frac{1}{2}{\rm Re}[ e_{z}(z)h_{y}(z)^{*} ]
\end{equation} 
given in Fig. \ref{CanPowPlotAll} for three solutions on each branch of solutions.
 Figure {\ref{CanPowPlotAll}}(a)  shows the power profile for three solutions on branch $1$ with $q/\ko=1.497+0.000106i$, $1.600+0.00150i$, and $2.569+0.0206i$ when $L_s=25$, $50$, and $95$ nm, respectively. The profiles  indicate that  the solutions for $L_s=25$ and $50$ nm represent surface waves   since the power is localized to the interfaces and decays away from the interface. However, the solution at $L_s=95$ nm represents a waveguide mode \cite{Yariv} since most of the power is guided by the lossy slab. The profiles also show a local maximum at $z=0$ inside the lossy slab indicating that this is a symmetric mode. This is the reason that branch $1$ is labeled symmetric. Similarly, Fig. {\ref{CanPowPlotAll}}(b) shows the power profiles for three solutions with $q/\ko=1.478+0.000215i$, $1.491+0.000847i$, and $2.364+0.00243i$  when $L_s=95$, $110$, and $190$ nm, respectively, on branch $2$. The power profiles in Fig. {\ref{CanPowPlotAll}}(b) indicate that the solutions at $L_s=95$ and $110$ nm are surface waves, and the solution at $L_s=190$ nm is a waveguide mode. Since, there is a zero at $z=0$, these solutions represent anti-symmetric modes. Similarly, the spatial power profiles in   {\ref{CanPowPlotAll}}(c)  for three solutions with $q/\ko=1.482+0.000894i$, $1.515+0.00315i$, and $1.805+0.0191i$  when $L_s=190$, $210$, and $250$ nm, respectively, on branch $3$, show that the same conclusion holds true, that is, the branch begins representing the Uller--Zenneck surface waves and converts to waveguide modes as $L_s$ increases. Furthermore, a local maximum at $z=0$ indicates that branch $3$  represents symmetric modes. Similarly, the power profiles in   Fig. {\ref{CanPowPlotAll}}(d) for $q/\ko=1.483+0.00152i$, $1.519+0.00482i$, and $1.783+0.0218i$ when $L_s=280$, $300$, and $340$ nm, respectively, on branch $4$, represent the anti-symmetric guided modes that transmutes from the Uller--Zenneck waves into waveguide modes.
 
 All the spatial profiles in Fig. \ref{CanPowPlotAll} confirm the hypothesis that each branch represents the Uller--Zenneck surface waves for smaller values of $L_s$ and represents waveguide modes for larger values of $L_s$. The important distinction between the surface waves and the waveguide modes is that the later quickly decays to zero outside of the lossy dielectric slab, whereas the former is localized to the interfaces and slowly decays away from the interface. Let us note that we could not find any quantitative criteria that can be used to mark the parts of each branch to distinguish the Uller--Zenneck waves from waveguide modes. The qualitative criteria, however, can be established based on the spatial power profiles. Therefore, the solutions represent Uller--Zenneck waves when most  of the power of the guided mode is localized to the interface,
and (ii) represent waveguide modes when most of the power is guided by the lossy slab and little power is present in the lossless partnering materials on both sides \cite{Yariv}. Furthermore, the solutions in between the clearly identifiable surface waves and the waveguide modes can be considered as hybrids of the both.
 
\subsection{Comparison with the SPP waves}
When the lossy dielectric material is replaced by a metal, the surface waves are SPP waves \cite{Raetherbook,SAM2007}. Furthermore, symmetric and anti-symmetric SPP waves are possible when the thickness of the slab is small. When the slab is thick, the anti-symmetric SPP-wave-mode transmutes into the symmetric mode and only symmetric SPP waves exist because the two interfaces decouple. Furthermore, no waveguide modes propagate. The Uller--Zenneck waves share one property with the SPP waves, that is, the Uller--Zenneck waves also come with either symmetric or anti-symmetric profiles; however, the possibility of waveguide modes in the lossy slab makes their evolution  with the increase in thickness different from that of the SPP waves. Both the symmetric and anti-symmetric Uller--Zenneck waves evolves into waveguide modes as the thickness increases. 

To see the symmetry and anti-symmetry of the Uller--Zenneck waves explicitly, let us proceed with the evaluation of  Eq. (\ref{dissp}) analytically, which leads to
\begin{equation}
\le\frac{\eps_+}{k_+}+\frac{\eps_-}{k_-}\ri-\le\frac{\eps_s}{k_s}+\frac{\eps_+\eps_-k_s}{k_+k_-\eps_s}\ri\tanh\le\frac{k_sL_s}{i}\ri=0\,,\label{dissp2}
\end{equation}  
after the determination of the determinant, where
\begin{equation}
k_\pm=\sqrt{\ko^2\mu_\pm\eps_\pm-q^2}\,,\quad k_s=\sqrt{\ko^2\mu_s\eps_s-q^2}\,.\nonumber
\end{equation}
When, $\eps_+=\eps_-=\eps_d$, $\mu_+=\mu_-=\mu_d$, Eq. (\ref{dissp2}) can be simplified to yield
\begin{equation}
\les\tanh\le\frac{k_sL_s}{2i}\ri+\frac{\eps_dk_s}{k_d\eps_s}\ris\les\tanh\le\frac{k_sL_s}{2i}\ri+\frac{\eps_sk_d}{k_s\eps_d}\ris=0
\end{equation}
that has two types of solutions:
\begin{eqnarray}
&&\tanh\le\frac{k_sL_s}{2i}\ri+\frac{\eps_sk_d}{k_s\eps_d}=0\,,\label{sol1}\\
&&\tanh\le\frac{k_sL_s}{2i}\ri+\frac{\eps_dk_s}{k_d\eps_s}=0\,.\label{sol2}
\end{eqnarray}
The dispersion equations (\ref{dissp2})--(\ref{sol2}) are in agreement with Refs. \cite{Raetherbook} and \cite{SAM2007} for the SPP waves guided by a metallic slab. Raether \cite{Raetherbook} derived it for SPP waves by finding the zeroes of the reflectance from a metallic strip, and Maier \cite{SAM2007} derived them by writing plane wave solutions in the three materials and implementing the boundary conditions.

The solutions of Eq. (\ref{sol1}) represent symmetric surface waves and those of Eq. (\ref{sol2}) represent anti-symmetric surface waves. We checked that the solutions presented in Fig. \ref{qvsLplots} on branches $1$ and $3$ are indeed solutions of Eq. (\ref{sol1}) and the solutions represented by branches $2$ and $4$ are solutions of Eq. (\ref{sol2}). Therefore, the Uller--Zenneck surface waves also come in either symmetric or anti-symmetric form; however, they do not transmute into one-type of solutions as is the case for SPP waves when the thickness $L_s$ increases. Instead, they evolve into waveguide modes as the thickness of the lossy slab increase.

 %%%%%%%%%%%%%%%%%%%%%%%%%%%%%%%%%%%%%%%
\begin{figure}
\includegraphics[width = 1\linewidth]{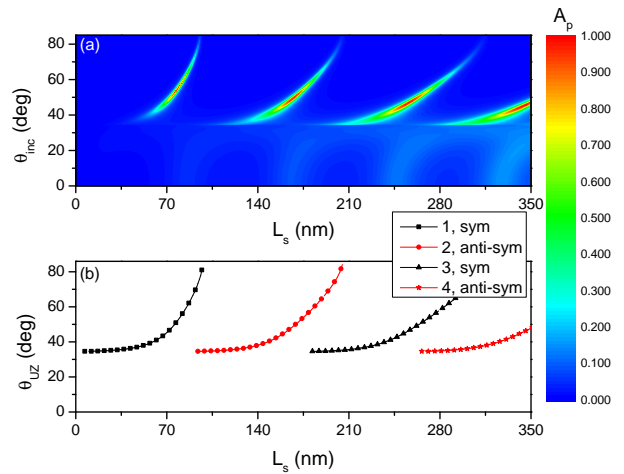}
\caption{{\bf Prism-coupled configuration:} (a) Absorptance $A_p$ as a function of the incidence angle $\theta_{inc}$ and the thickness of the lossy dielectric material $L_{s}$ when $\lambdao=633$ nm, $L_D=100$ nm, $n_p=2.6$, $\varepsilon_{\pm} = 2.179$, $\varepsilon_{s} = 15.072 + 0.1521i$, and $\mu_{\pm} = \mu_{s} = 1$. (b) The incidence angle $\theta_{\rm UZ}$ where the canonical boundary-value problem predicts the excitation of a Uller--Zenneck wave, computed using Eq. (\ref{thetauz}).} 
\label{Abs}
\end{figure}
%%%%%%%%%%%%%%%%%%%%%%%%%%%%%%%%%%%%%%%

%%%%%%%%%%%%%%%%%%%%%%%%%%%%%%%%%%%%
\begin{figure*}
\includegraphics[width = 5.5in]{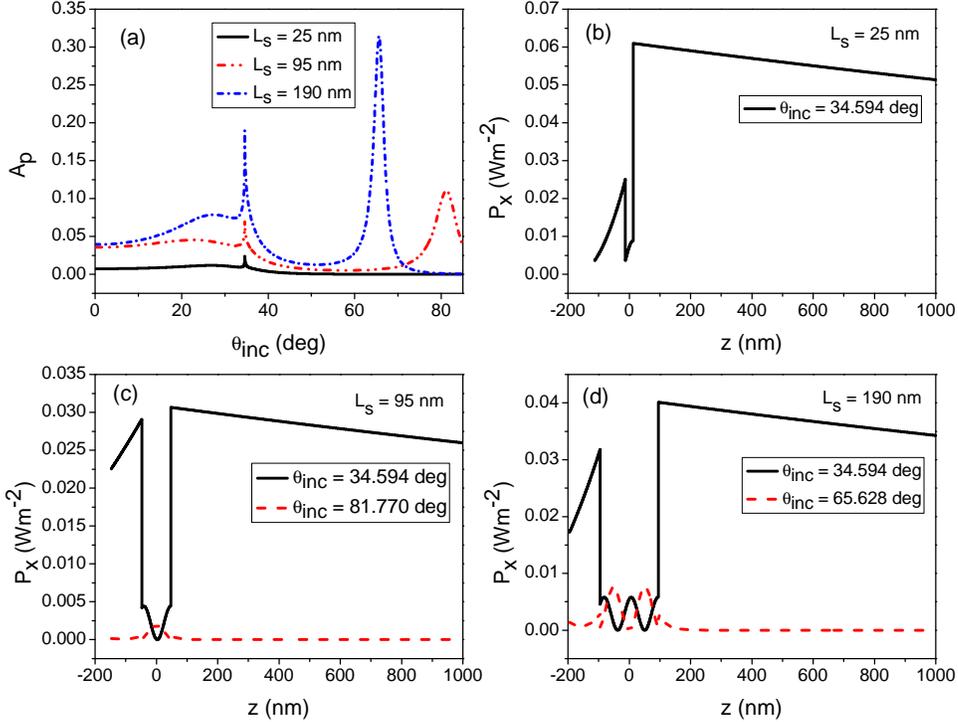}
\caption{{\bf Prism-coupled configuration: } (a) Absorptance $A_p$ as function of $\theta_{inc}$ when $L_s=25$, $95$, and $190$ nm.  (b-d) Variation of the $x$-component of the time-averaged Poynting vector $P_{x}$ as a function of $z$ for those incidence angles where a peak exist in the plots of $A_p$ in part (a). For the computation, $a_p=1$ Vm$^{-1}$ was set. Other parameters are the same as in Fig. \ref{Abs}(a). } 
\label{Powpr}
\end{figure*}
%%%%%%%%%%%%%%%%%%%%%%%%%%%%%%%%%%

\subsection{Excitation}
Let us now discuss the excitation of the Uller--Zenneck waves in the prism coupled configuration. For this purpose, the absoprtance $A_p$ for $p$-polarized incident plane wave is shown in Fig. \ref{Abs}(a) when a rutile prism ($n_p=2.6$) is used with a finitely thick layer of silica  between the prism and silicon, whereas the upper half-space is assumed to be occupied by silica. The figure clearly shows the sharp absorptance bands showing the excitation of either the Uller--Zenneck surface waves or the waveguide modes. For comparison with the canonical boundary-value problem, the incidence angle where a guided mode should be excited was computed as
 \begin{equation}
\theta_{UZ} = \sin^{-1}\Bigg[\frac{{\rm Re}(q)}{n_{p}k_{0}}\Bigg]\label{thetauz}
\end{equation}
 and is shown in Fig. \ref{Abs}(b). A comparison of Figs. \ref{Abs}(a) and (b) shows that the prediction of the canonical problem matches excellently with the results of the prism-coupled configuration.

The plot of absorptance $A_p$ is shown in Fig. \ref{Powpr}(a) as a function of the incidence angle when $L_s=25$, $95$,  and $190$ nm. The figure clearly shows an absorptance peak at $\theta_{inc}=34.594^\circ$ independent of the value of $L_s$. However, as $L_s$ increases from $25$ nm, other peaks also appear at different values of the incidence angle. The fixed peak at $\theta_{inc}=34.594^\circ$ represents the Uller--Zenneck surface wave, whereas other peaks represent  the waveguide modes. This is also evident from the spatial profiles of the $x$-component of the time-averaged Poynting vector $P_x$ in Figs. \ref{Powpr}(b), (c), and (d). These figures show that the peaks at $\theta_{inc}=34.594^\circ$ represent surface waves since the power profile is localized to the interfaces and decay away from them, whereas the the peaks at other incidence angles represent waveguide modes since $P_x$ is non-zero inside the lossy dielectric material and quickly decays to zero outside of it. Therefore, the Uller--Zenneck waves can be excited in the prism-coupled configuration, all with the planar interfaces.

\section{Concluding Remarks}\label{conc}
A canonical boundary-value problem was set up and solved to find the wavenumbers of the Uller--Zenneck surface waves and the waveguide modes guided by a slab of a lossy dielectric material with lossless dielectric materials on both sides. The solution of the canonical problem showed that Uller--Zenneck surface waves are guided by  the slab of the lossy dielectric material and that the surface waves are localized to  the interfaces and decay away from the interfaces. The Uller--Zenneck surface waves have either symmetric or anti-symmetric spatial power profiles. Furthermore, the Uller--Zenneck waves transmute into the waveguide modes as the thickness of the slab of the lossy dielectric increases in contrast to the surface plasmon-polariton (SPP) waves guided by a metallic slab that do not support waveguide modes.  The waveguide modes were confined to within the slab and decayed quickly out of the slab. The number of waveguide modes increased as the thickness of the slab increased.

A prism-coupled configuration was used to elucidate the excitation of the Uller--Zenneck waves and it was observed that a Uller--Zenneck wave is  excited for any thickness of the slab of the lossy dielectric material. Also, the angle of incidence of the Uller--Zenneck waves was independent of the thickness of the slab of lossy material. The waveguide modes exist only when the slab is sufficiently thick and their number increases as the thickness of the slab increases. Furthermore, the angle of incidence of waveguide modes changed with the change in the thickness of the slab.

This proposed prism-coupled configuration will help usher the exploitation of the Uller--Zenneck surface waves since we showed that it can be excited with the planar interfaces between dielectric materials, with at least one being lossy.  The   Uller--Zenneck waves have potential for optical sensing, just like SPP waves \cite{JH2006,SAM2007}; however, without the need of a metallic film. Also, the Uller--Zenneck waves can  lead to new applications of the surface waves as it does not require a metal to excite in contrast to the surface plasmon-polariton (SPP) waves that need metal \cite{all}.

\end{document}